\documentclass[a4paper,12pt]{article}

\usepackage{newlfont}
\usepackage{amsfonts}
\usepackage{amsthm}
\usepackage{amssymb}
\usepackage{amsmath}
\usepackage{latexsym}
\usepackage[all]{xy}

-\headheight\advance\topmargin by -\headsep

\numberwithin{equation}{section} \allowdisplaybreaks

\newtheorem{proposition}{Proposition}[section]

\theoremstyle{definition}

\begin{document}
\font\black=cmbx10 \font\sblack=cmbx7 \font\ssblack=cmbx5
\font\blackital=cmmib10  \skewchar\blackital='177
\font\sblackital=cmmib7 \skewchar\sblackital='177
\font\ssblackital=cmmib5 \skewchar\ssblackital='177
\font\sanss=cmss10 \font\ssanss=cmss8 
\font\sssanss=cmss8 scaled 600 \font\blackboard=msbm10
\font\sblackboard=msbm7 \font\ssblackboard=msbm5
\font\caligr=eusm10 \font\scaligr=eusm7 \font\sscaligr=eusm5
\font\blackcal=eusb10 \font\fraktur=eufm10 \font\sfraktur=eufm7
\font\ssfraktur=eufm5 \font\blackfrak=eufb10

\font\bsymb=cmsy10 scaled\magstep2
\def\all#1{\setbox0=\hbox{\lower1.5pt\hbox{\bsymb
       \char"38}}\setbox1=\hbox{$_{#1}$} \box0\lower2pt\box1\;}
\def\exi#1{\setbox0=\hbox{\lower1.5pt\hbox{\bsymb \char"39}}
       \setbox1=\hbox{$_{#1}$} \box0\lower2pt\box1\;}

\def\mi#1{{\fam1\relax#1}}
\def\tx#1{{\fam0\relax#1}}

\newfam\bifam
\textfont\bifam=\blackital \scriptfont\bifam=\sblackital
\scriptscriptfont\bifam=\ssblackital
\def\bi#1{{\fam\bifam\relax#1}}

\newfam\blfam
\textfont\blfam=\black \scriptfont\blfam=\sblack
\scriptscriptfont\blfam=\ssblack
\def\rbl#1{{\fam\blfam\relax#1}}

\newfam\bbfam
\textfont\bbfam=\blackboard \scriptfont\bbfam=\sblackboard
\scriptscriptfont\bbfam=\ssblackboard
\def\bb#1{{\fam\bbfam\relax#1}}

\newfam\ssfam
\textfont\ssfam=\sanss \scriptfont\ssfam=\ssanss
\scriptscriptfont\ssfam=\sssanss
\def\sss#1{{\fam\ssfam\relax#1}}

\newfam\clfam
\textfont\clfam=\caligr \scriptfont\clfam=\scaligr
\scriptscriptfont\clfam=\sscaligr
\def\cl#1{{\fam\clfam\relax#1}}

\newfam\frfam
\textfont\frfam=\fraktur \scriptfont\frfam=\sfraktur
\scriptscriptfont\frfam=\ssfraktur
\def\fr#1{{\fam\frfam\relax#1}}

\def\cb#1{\hbox{$\fam\gpfam\relax#1\textfont\gpfam=\blackcal$}}

\def\hpb#1{\setbox0=\hbox{${#1}$}
    \copy0 \kern-\wd0 \kern.2pt \box0}
\def\vpb#1{\setbox0=\hbox{${#1}$}
    \copy0 \kern-\wd0 \raise.08pt \box0}

\def\pmb#1{\setbox0\hbox{${#1}$} \copy0 \kern-\wd0 \kern.2pt \box0}
\def\pmbb#1{\setbox0\hbox{${#1}$} \copy0 \kern-\wd0
      \kern.2pt \copy0 \kern-\wd0 \kern.2pt \box0}
\def\pmbbb#1{\setbox0\hbox{${#1}$} \copy0 \kern-\wd0
      \kern.2pt \copy0 \kern-\wd0 \kern.2pt
    \copy0 \kern-\wd0 \kern.2pt \box0}
\def\pmxb#1{\setbox0\hbox{${#1}$} \copy0 \kern-\wd0
      \kern.2pt \copy0 \kern-\wd0 \kern.2pt
      \copy0 \kern-\wd0 \kern.2pt \copy0 \kern-\wd0 \kern.2pt \box0}
\def\pmxbb#1{\setbox0\hbox{${#1}$} \copy0 \kern-\wd0 \kern.2pt
      \copy0 \kern-\wd0 \kern.2pt
      \copy0 \kern-\wd0 \kern.2pt \copy0 \kern-\wd0 \kern.2pt
      \copy0 \kern-\wd0 \kern.2pt \box0}

\def\cdotss{\mathinner{\cdotp\cdotp\cdotp\cdotp\cdotp\cdotp\cdotp
        \cdotp\cdotp\cdotp\cdotp\cdotp\cdotp\cdotp\cdotp\cdotp\cdotp
        \cdotp\cdotp\cdotp\cdotp\cdotp\cdotp\cdotp\cdotp\cdotp\cdotp
        \cdotp\cdotp\cdotp\cdotp\cdotp\cdotp\cdotp\cdotp\cdotp\cdotp}}

\font\frak=eufm10 scaled\magstep1 \font\fak=eufm10 scaled\magstep2
\font\fk=eufm10 scaled\magstep3 \font\scriptfrak=eufm10
\font\tenfrak=eufm10


\mathchardef\za="710B  
\mathchardef\zb="710C  
\mathchardef\zg="710D  
\mathchardef\zd="710E  
\mathchardef\zve="710F 
\mathchardef\zz="7110  
\mathchardef\zh="7111  
\mathchardef\zvy="7112 
\mathchardef\zi="7113  
\mathchardef\zk="7114  
\mathchardef\zl="7115  
\mathchardef\zm="7116  
\mathchardef\zn="7117  
\mathchardef\zx="7118  
\mathchardef\zp="7119  
\mathchardef\zr="711A  
\mathchardef\zs="711B  
\mathchardef\zt="711C  
\mathchardef\zu="711D  
\mathchardef\zvf="711E 
\mathchardef\zq="711F  
\mathchardef\zc="7120  
\mathchardef\zw="7121  
\mathchardef\ze="7122  
\mathchardef\zy="7123  
\mathchardef\zf="7124  
\mathchardef\zvr="7125 
\mathchardef\zvs="7126 
\mathchardef\zf="7127  
\mathchardef\zG="7000  
\mathchardef\zD="7001  
\mathchardef\zY="7002  
\mathchardef\zL="7003  
\mathchardef\zX="7004  
\mathchardef\zP="7005  
\mathchardef\zS="7006  
\mathchardef\zU="7007  
\mathchardef\zF="7008  
\mathchardef\zW="700A  

\newcommand{\be}{\begin{equation}}
\newcommand{\ee}{\end{equation}}
\newcommand{\ra}{\rightarrow}
\newcommand{\lra}{\longrightarrow}
\newcommand{\bea}{\begin{eqnarray}}
\newcommand{\eea}{\end{eqnarray}}
\newcommand{\beas}{\begin{eqnarray*}}
\newcommand{\eeas}{\end{eqnarray*}}
\def\*{{\textstyle *}}
\newcommand{\R}{{\mathbb R}}
\newcommand{\T}{{\mathbb T}}
\newcommand{\C}{{\mathbb C}}
\newcommand{\unit}{{\mathbf 1}}
\newcommand{\SL}{SL(2,\C)}
\newcommand{\Sl}{sl(2,\C)}
\newcommand{\SU}{SU(2)}
\newcommand{\su}{su(2)}
\def\ssT{\sss T}
\newcommand{\G}{{\goth g}}
\newcommand{\D}{{\rm d}}
\newcommand{\Df}{{\rm d}^\zF}
\newcommand{\de}{\,{\stackrel{\rm def}{=}}\,}
\newcommand{\we}{\wedge}
\newcommand{\nn}{\nonumber}
\newcommand{\ot}{\otimes}
\newcommand{\s}{{\textstyle *}}
\newcommand{\ts}{T^\s}
\newcommand{\oX}{\stackrel{o}{X}}
\newcommand{\oD}{\stackrel{o}{D}}
\newcommand{\obD}{\stackrel{o}{\bD}}
\newcommand{\pa}{\partial}
\newcommand{\ti}{\times}
\newcommand{\A}{{\cal A}}
\newcommand{\Li}{{\cal L}}
\newcommand{\ka}{\mathbb{K}}
\newcommand{\find}{\mid}
\newcommand{\ad}{{\rm ad}}
\newcommand{\rS}{]^{SN}}
\newcommand{\rb}{\}_P}
\newcommand{\p}{{\sf P}}
\newcommand{\h}{{\sf H}}
\def\d{\sf{d}}
\newcommand{\X}{{\cal X}}
\newcommand{\I}{\,{\rm i}\,}
\newcommand{\rB}{]_P}
\newcommand{\Ll}{{\pounds}}
\def\lna{\lbrack\! \lbrack}
\def\rna{\rbrack\! \rbrack}
\def\rnaf{\rbrack\! \rbrack_\zF}
\def\rnah{\rbrack\! \rbrack\,\hat{}}
\def\lbo{{\lbrack\!\!\lbrack}}
\def\rbo{{\rbrack\!\!\rbrack}}
\def\lan{\langle}
\def\ran{\rangle}
\def\zT{{\cal T}}
\def\tU{\tilde U}
\def\ati{{\stackrel{a}{\times}}}
\def\sti{{\stackrel{sv}{\times}}}
\def\aot{{\stackrel{a}{\ot}}}
\def\sati{{\stackrel{sa}{\times}}}
\def\saop{{\stackrel{sa}{\op}}}
\def\bwa{{\stackrel{a}{\bigwedge}}}
\def\svop{{\stackrel{sv}{\oplus}}}
\def\saot{{\stackrel{sa}{\otimes}}}
\def\cti{{\stackrel{cv}{\times}}}
\def\cop{{\stackrel{cv}{\oplus}}}
\def\dra{{\stackrel{\xd}{\ra}}}
\def\bdra{{\stackrel{\bd}{\ra}}}
\def\bAff{\mathbf{Aff}}
\def\Aff{\sss{Aff}}
\def\bHom{\mathbf{Hom}}
\def\Hom{\sss{Hom}}
\def\bt{{\boxtimes}}
\def\sot{{\stackrel{sa}{\ot}}}
\def\bp{{\boxplus}}
\def\op{\oplus}
\def\bwak{{\stackrel{a}{\bigwedge}\!{}^k}}
\def\aop{{\stackrel{a}{\oplus}}}
\def\ix{\operatorname{i}}
\def\V{{\cal V}}
\def\cD{{\cal D}}
\def\cC{{\cal C}}
\def\cE{{\cal E}}
\def\cL{{\cal L}}
\def\cN{{\cal N}}
\def\cR{{\cal R}}
\def\cJ{{\cal J}}
\def\cT{{\cal T}}
\def\cH{{\cal H}}
\def\bA{\mathbf{A}}
\def\bI{\mathbf{I}}
\def\wh{\widehat}
\def\wt{\widetilde}
\def\ol{\overline}
\def\ul{\underline}
\def\Sec{\sss{Sec}}
\def\Lin{\sss{Lin}}
\def\ader{\sss{ADer}}
\def\ado{\sss{ADO^1}}
\def\adoo{\sss{ADO^0}}
\def\AS{\sss{AS}}
\def\bAS{\sss{AS}}
\def\bLS{\sss{LS}}
\def\bAP{\sss{AV}}
\def\bLP{\sss{LP}}
\def\AP{\sss{AP}}
\def\LP{\sss{LP}}
\def\LS{\sss{LS}}
\def\Z{\mathbf{Z}}
\def\Y{\mathbf{Y}}
\def\oZ{\overline{\bZ}}
\def\oA{\overline{\bA}}
\def\cim{{C^\infty(M)}}
\def\de{{\cal D}^1}
\def\la{\langle}
\def\ran{\rangle}
\def\by{{\bi y}}
\def\bs{{\bi s}}
\def\bc{{\bi c}}
\def\bd{{\bi d}}
\def\bh{{\bi h}}
\def\bD{{\bi D}}
\def\bY{{\bi Y}}
\def\bX{{\bi X}}
\def\bL{{\bi L}}
\def\bV{{\bi V}}
\def\bW{{\bi W}}
\def\bS{{\bi S}}
\def\bT{{\bi T}}
\def\bC{{\bi C}}
\def\bE{{\bi E}}
\def\bF{{\bi F}}
\def\bP{{\bi P}}
\def\bp{{\bi p}}
\def\bz{{\bi z}}
\def\bZ{{\bi Z}}
\def\bq{{\bi q}}
\def\bQ{{\bi Q}}
\def\bx{{\bi x}}

\def\sA{{\sss A}}
\def\sC{{\sss C}}
\def\sD{{\sss D}}
\def\sd{{\sss d}}
\def\sG{{\sss G}}
\def\sH{{\sss H}}
\def\sI{{\sss I}}
\def\sJ{{\sss J}}
\def\sK{{\sss K}}
\def\sL{{\sss L}}
\def\sO{{\sss O}}
\def\sP{{\sss P}}
\def\sPh{{\sss P\sss h}}
\def\sT{{\sss T}}
\def\sV{{\sss V}}
\def\sR{{\sss R}}
\def\sS{{\sss S}}
\def\sE{{\sss E}}
\def\sF{{\sss F}}
\def\st{{\sss t}}
\def\sg{{\sss g}}
\def\sx{{\sss x}}
\def\sv{{\sss v}}
\def\sw{{\sss w}}
\def\sQ{{\sss Q}}
\def\sj{{\sss j}}
\def\sq{{\sss q}}
\def\xa{\tx{a}}
\def\xc{\tx{c}}
\def\xd{\tx{d}}
\def\xi{\tx{i}}
\def\xD{\tx{D}}
\def\xV{\tx{V}}
\def\xF{\tx{F}}

\title{AV-differential geometry and calculus of variations\thanks{Research
supported by the Polish Ministry of Scientific Research and
Information Technology under the grant No. 2 P03A 036 25.}}

        \author{
        Katarzyna  Grabowska, Pawe\l\ Urba\'nski\\
        \\
        {\it Physics Department}\\
                {\it University of Warsaw}          
                }


\maketitle

\thanks{Supported by KBN, Grant 2PO3A 041 18, Grant 2PO3A 036 25}

\begin{abstract}
The calculus of variations for lagrangians which are not functions
on the tangent bundle, but sections certain affine bundles is
developed. We follow a general approach to variational principles
which admits boundary terms of variations.

\bigskip\noindent
\textit{MSC 2000: 70G45, 70H03, 70H05}

\medskip\noindent
\textit{Key words: affine spaces, Lagrangian formalism, Euler-Lagrange equation}

\end{abstract}

\section{Introduction}

It is already commonly accepted that the gauge independent
lagrangian for relativistic charged particle is not a function,
but a section of a bundle of affine lines over the tangent bundle
of the space-time manifold \cite{TU}. Also frame-independent
lagrangian for Newtonian particle is not a function, but a section
of an affine bundle \cite{GU}. In \cite{GGU,U} we have shown that
the proper geometric tools for such a frame-independent
formulation of Lagrangian systems are provided by the {\it
geometry of affine values (AV- differential geometry)}. We call
the geometry of affine values the differential geometry which is
built using sections of a principal $\R$-bundle over the manifold
instead of functions on the manifold. Since the bundle we use is
equipped with the fiber action of the group $(\R,+)$, we can add
reals to elements of fibres and real functions to sections, but
there is no distinguished "zero section".

In the present paper we apply these tools to develop the calculus
of variations for lagrangians with affine values. First, we
recognize the affine space of values of the action functional.
Then, we find affine analogues of constructions, which lead to the
proper representation of the differential of the action
functional. Finally, we formulate variational principles for
lagrangians with affine values.
Note that the affine analogues of the Euler-Lagrange equations
have been obtained purely geometrically in \cite{IMPS,GGU1}.

\section{Bundles of affine values}

  An affine bundle $\zeta:\Z\rightarrow M$, modelled on the trivial vector bundle
  $M \times \R \rightarrow M$, will be called a {\it  bundle of affine values}
  (shortly, an {\it AV-bundle}). We can say equivalently that an AV-bundle
  is a principal bundle  with the structural group $(\R , +)$.

An  AV-bundle is canonically associated with any special affine
bundle $\bA=(A,v_\bA)$, i.e. an affine bundle $A$ with a
distinguished nowhere-vanishing section $v_\bA$ of the model
vector bundle $\sV(A)$. The free $\R$-action induced from
translations in the direction of $-v_\bA$ makes $A$ an into an
$(\R , +)$-principal bundle over $A/\la v_\bA\ran$. This AV-bundle
we will denote $\bAP(\bA)$.

An {\it affine covector} is  an equivalence class of the relation
defined in the set of pairs of $(m,\zf)$, where $m\in M$ and $\zf$
is a section of $\Z$. We say that $(m,\zf)$, $(m',\zf')$ are {\it
equivalent} if $m=m'$ and $\xd(\zf-\zf')(m)=0$, where we have
identified the difference of sections of $\Z$ with a function on
$M$. The equivalence class of $(m,\zf)$ is denoted by $\xd\zf(m)$.
An affine analogue of the cotangent bundle is the union of all
affine covectors. It is an affine bundle modelled on $T^\ast M$
and equipped with a canonical symplectic form. We denote it by
$\sP\Z$ and we call it the {\it phase bundle} for $\Z$.

Since in the space of affine 1-forms (sections of $\sP\Z
\rightarrow M$) we have a distinguished family of exact forms
$\xd\zf$ and the de Rham differential $\xd(\zs-\xd\zf)$ does not
depend on $\zf$, there is a well-defined affine de Rham
differential from affine 1-forms into affine 2-forms, which are
ordinary 2-forms on $M$. The de Rham complex for $\Z$ can be then
continued as in the standard case.

\section{Other examples of affine constructions}
\label{afcon}

   In the following, we need two simple constructions concerning affine spaces and affine forms.

    Let $A,B$ be two affine spaces with the same model vector space $V$. In the product
    $A \times B$, we define an equivalence relation. Two pairs $(a,b)$ and $(a',b')$
    are equivalent if $a - a' = b' - b$. The equivalence class  of a pair $(a,b)$ is,
    by definition, the {\it sum} of $a$ and $b$. We denote it by $a\boxplus b$. The family
    of all sums is, obviously, an affine space modelled on $V$ with
                    $$a\boxplus b  -  a'\boxplus b' = (a-a') + (b-b')$$
and we denote it by $A \boxplus B$.  Similarly, we define $A  \boxminus B$. 

   The same method can be used to define an affine space of values of integrals of affine
forms along an oriented 1-dimensional cell in $M$. The value of
the integral is the equivalence class of pairs $(\zs, I)$, where
$\zs$ is an affine 1-form, i.e. a section of the phase bundle $\sP
\bZ$, and $I$ is a cell in $M$. Two pairs $(\zs, I)$ and
$(\zs',I')$ are equivalent if $I=I'$, and
    $$\int_I (\zs - \zs') = 0$$
The equivalence class of a pair $(\zs, I)$ will be denoted by
$\int _I\zs$. It is obvious that the set of values of integrals
along $I$ is an affine space with the model vector space $\R$.
The linear part of $\int_I$ is the standard integral of 1-forms.

    \begin{proposition}
Let $\zg \colon [a,b] \rightarrow M$ be a parameterization of a
cell $I$. Then the space of values of the integral is canonically
isomorphic to $Z_{\zg(b)} \boxminus Z_{\zg(a)}$.
    \end{proposition}
    \begin{proof}
For $\zs = \xd \zf$ we put $J(\zs) = \zf (\zg(b)) \boxminus \zf
(\zg(a))  \in Z_{\zg(b)} \boxminus Z_{\zg(a)}$. It is obvious that
the mapping $J$ is well-defined and surjective. Moreover
        $$J(\xd(\zf +f)) = J(\xd\zf) + (f(\zg(b)) -  f(\zg(b))) = J(\xd\zf) + \int_I \xd f  ,$$
    i.e. the linear parts of $J$ and $\int_I$ coincide on exact forms.
    It follows that the mapping
            $$ \int_I \xd\zf \mapsto J(\xd\zf)$$
is well defined and gives canonical isomorphism of $\in Z_{\zg(b)}
\boxminus Z_{\zg(a)}$ and the AV-space for $\int_I$.
    \end{proof}

\section{Special vector spaces an duality}
\label{dual} Every finite-dimensional vector space $V$ can be
considered as the space of linear functions on its dual $V^\*$,
i.e. the space of linear sections of the trivial bundle $V^\*
\times \R$. This bundle is an example of an AV-bundle associated
with a special vector space, namely the space $V\times \R$ with
the distinguished non-zero vector $(0,1)$. Also elements of an
affine space $A$ with the model vector space $V$ can be
interpreted as linear sections of an AV-bundle $\zt \colon A^\dag
\rightarrow V^\*$. The canonical choice for $A^\dag$ is the
special vector space of affine functions on $A$. The projection
$\zt$ associates with an affine function its linear part.

\begin{proposition}\label{pr1} For an AV-bundle
$\Z=(Z,v_\bZ)$, we have a canonical isomorphism of special vector
spaces $((\sP_m\Z) ^\dag, 1)$ and $(\sT_z Z, v_0)$, where $m =
\zeta (z)$ and $v_0$ is the fundamental vector for the group
action on $Z$.
    \end{proposition}
    \begin{proof}
    Let  $a = \xd_m \zf$ with $\zf(m) =z$.  The mapping

\begin{equation}\label{d1}
      a \colon \sT_m M \rightarrow \sT_z Z \colon v \mapsto \sT \zf(v)
\end{equation}
is a linear section of the AV-bundle, associated to special vector
space $(\sT_z Z, v_0)$. By duality, we get canonical isomorphism
of $(\sP_m\Z)^\dag$ and  $(\sT_z Z, v_0)$.
    \end{proof}

The group action on $Z$ gives isomorphisms of tangent spaces along
fibers of $\zeta$. It follows that the bundle $(\sP \Z)^\dag$ can
be identified with $\wt{\sT} \Z = \sT \Z /\R$. We conclude that
there is a one-to one correspondence between linear sections of
$\wt{\sT}\zeta \colon  \wt{\sT}\Z \rightarrow \sT M $ and affine
one-forms. In the following we use this isomorphism to identify
affine covectors in terms of its affine values.

\subsection{The action functional}
\label{inhom}

It is known that the lagrangians for relativistic charged
particles, lagrangians in Newtonian mechanics, etc., are not
functions but sections of an AV-bundle of the form $\wt{\sT}\zeta
\colon \wt{\sT}\Z \rightarrow \sT M $ for certain bundle $\zeta
\colon Z \rightarrow M$. The action for such lagrangian can be
defined in the following way. First, we  observe that by
Proposition \ref{pr1} every affine 1-form, i.e. a section $\zs$ of
the phase bundle $\sP\Z$, defines a section of  $\wt{\sT}\zeta$.
We denote it by $\xi^a_\sT \zf$.
    Let $\zg \colon [a,b] \rightarrow M$. We put
$$ \int_a^b  \xi^a_\sT\zf \circ\dot{\zg}: =
\int_{\zg([a,b])} \zf \in Z_{\zg(b)} \boxminus Z_{\zg(a)},$$
and for any lagrangian $\zl \colon \sT M \rightarrow \wt{\sT} \Z$,

$$\int_a^b \zl \circ \dot{\zg} = \int_a^b (\zl \circ \dot{\zg} - \xi^a_\sT\zf \circ\dot{\zg})
+ \int_a^b  \xi^a_\sT\zf \circ\dot{\zg} \ \ .$$

The action for a lagrangian $\zl$ can be defined in a different
(but equivalent) way. Let $\zg^\* \Z$ be the pull-back of $\Z$
with respect to the curve $\zg$. We have mappings $\wt{\zg} \colon
\zg^\* \Z \rightarrow \Z$ and
$$\wt{\sT} \wt{\zg} \colon \wt{\sT} \zg^\* \Z \rightarrow  \wt{\sT}\Z ,$$
defined in an obvious way. It is clear that the image of $\zl
\circ \dot{\zg}$ belongs to the image of $\wt{\sT} \wt{\zg}$.
Consequently, there is a unique section
    $$\zl_\zg \colon [a,b] \rightarrow \wt{\sT} (\zg^\* \Z) $$
    such that
    $$   (\wt{\sT} \wt{\zg})\circ \zl_\zg = \zl \circ \dot{\zg} .$$
The section $\zl_\zg$ defines an $\R$-invariant vector field on
$\zg^\* \Z$ and its family of integral curves is also
$\R$-invariant. It follows that for each integral curve
$\wt{\zl_\zg}\colon [a,b] \rightarrow \zg^\* \Z$ we can define the
affine number $\wt{\zl_\zg} (b) \boxminus  \wt{\zl_\zg} (a)$ which
does not depend on the choice of the curve.
 \begin{proposition}\label{pr2}
    For each curve $\zg \colon [a,b] \rightarrow M$, we have
$$\wt{\zl_\zg} (b) \boxminus  \wt{\zl_\zg} (a) = \int_a^b \zl \circ \dot{\zg} \ \ .$$
    \end{proposition}

\section{Operations on affine 1-forms}
\subsection{Pull-back of an affine 1-form} \label{pull-back}

Let $g \colon N \rightarrow M$ be a differentiable mapping and let
$\zp \colon P \rightarrow M$ be an affine bundle, modelled on
$\sT^\* M$. The AV-bundle for $P$ is $\zeta_P \colon P^\dag
\rightarrow \sT M$. The bundle $\zp_g \colon P_g \rightarrow N$ we
define as the dual to the pull-back of the bundle of affine values
for $P$, with respect to the tangent mapping $\sT g$:

\begin{equation}
    (P_g)^\dag =(\sT g)^\* P^\dag .
\end{equation}

    The pull-back $g^\*\zs$ of a section $\zs$ of $\zp$ is defined by
     \begin{equation}
    \xi^a_\sT(g^\* \zs)=(\sT g)^\* \xi^a_\sT (\zs) .
\end{equation}
If the bundle $P$ is the phase bundle for an AV-bundle $\Z$, then
we have the obvious isomorphism
        \begin{equation}
    (\sP \Z)_g = \sP (g^\*\Z)
\end{equation}
    and, as in the standard case,

    \begin{equation}
    g^\* \xd\zf =  \xd(g^\* (\zf)),
\end{equation}
    i.e. the pull-back commutes with the affine exterior differential.

\subsection{The total derivative $\xd^a_\sT$}

Derivations $\xd_\sT$ and  $\xi_F$ play fundamental role in the
geometric integration by parts - the central point in the calculus
of variations.  Both, $\xd_\sT$ and  $\xi_F$ are derivations
defined on the algebra of differential forms on a manifold $N$,
with values in differential forms on $\sT N$. Here, we define
their  affine counter-parts, but for affine 1-forms only.

In the standard geometry, the total derivative $\xd_\sT$ can be
defined in two  equivalent ways. First,  as the commutator
    $$\xd_\sT = \xi_\sT \xd + \xd \xi_\sT \ , $$
where $\xi_\sT$ is the derivation given by the identity mapping on
$\sT N$, interpreted as a vector-valued function on $\sT N$, with
values in $\sT N$. Second, by the formula

\begin{equation}\label{dt1}
  \xi_\sT (\xd_\sT \zw) = (\xd_\sT \xi_\sT \zw)\circ \zk_N \ ,
\end{equation}
where $\zw$ is a 1-form, and $\zk_N\colon \sT\sT N \rightarrow \sT\sT N$ is the canonical flip.

Both formulae  have their counterparts in the affine case. Let $\Y $ be an AV-bundle over $N$.
      The affine total derivative can be defined by the formula

\begin{equation}\label{dt}
  \xd^a_\sT (\zs) = \xi_\sT \xd \zs + \xd \xi^a_\sT \zs
\end{equation}
for a section $\zs$ of $\sP \Z$. Since $\xd \zs$ is an ordinary
2-form, the first term in \ref{dt} is a 1-form on $\sT N$. The
second term  is a section of $\sP \wt{\sT} \Y $  and the
corresponding dual AV-bundle is $ \wt{\sT}\wt{\sT} \Y$. We
conclude that the AV-bundle for $\xd_\sT$ is also  $
\wt{\sT}\wt{\sT} \Y$.

Another definition of the total derivative is by a formula,
analogous to \ref{dt1}. First, we observe that the canonical flip
$\zk_Y \colon \sT\sT Y \rightarrow \sT \sT Y$ reduces to an
isomorphism

    $$ \wt{\zk}_Y \colon \wt{\sT}\wt{\sT} \Y \rightarrow \wt{\sT}\wt{\sT} \Y \ . $$
   \begin{proposition}\label{pr3}
   Let $\zs$ be an affine 1-form on $N$, i.e. a section of $\sP \Y$. then
   \begin{equation}\label{dt2}
   \xi^a_\sT (\xd_\sT \zs) = \wt{\zk}_Y \circ(\xi^a_\sT \xd ( \xi^a_\sT \zs))\circ \zk_N \
\end{equation}
    \end{proposition}
    \begin{proof}
The linear parts of \ref{dt} and \ref{dt2} coincide, so it is
enough to compare these formulae for $\zs = \xd \zf$, where $\zf$
is a section of the AV-bundle $\Y$. The formula  \ref{dt} gives in
this case
    $$ \xd^a_\sT \xd \zf = \xd \xi^a_\sT \xd \zf $$
    and the corresponding section of the AV-bundle

\begin{equation}\label{pr4}
  \xi^a_\sT (\xd_\sT \xd\zf)= \xi_\sT^a \xd \xi^a_\sT \xd \zf \ .
\end{equation}

It follows from  \ref{d1} that  $\xi^a_\sT \xd ( \xi^a_\sT \xd
\zf)$ is reduced $\sT \sT \zf$, i.e. $\wt{\sT} \wt{\sT} \zf$. The
canonical flip is an equivalence of functors, hence
$$\wt{\zk}_Y \circ (\wt{\sT} \wt{\sT} \zf) \circ \zk_N = \wt{\sT} \wt{\sT} \zf.$$
We obtain
   \begin{equation}\label{pr5}
\xi^a_\sT (\xd_\sT \xd\zf)= \xi_\sT^a \xd \xi^a_\sT \xd \zf  =
\wt{\zk}_Y \circ(\xi^a_\sT \xd ( \xi^a_\sT \xd\zf))\circ \zk_N \ .
\end{equation}
    \end {proof}
   \subsection{The operation $\xi^a_F$}
$\xi_F$ is a derivation in the algebra of forms on $\sT N$ given
by the vertical endomorphism $F \colon \sT\sT N \rightarrow \sT\sT
N$, interpreted as vector-valued 1-form. For a standard  1-form
$\zw$ on $\sT N$ we have $\xi_F \zw (w) = \zw (F w)$. The same
formula we use for affine 1-forms - sections of $\sP\wt{\sT} \Y$.
Since the AV-bundle for $\sP\wt{\sT} \Y$ is $ \wt{\sT} \wt{\sT}
\Y$, the AV-bundle for $\xi^a_\sT \zs$ is the pull-back bundle
$F^\* \wt{\sT} \wt{\sT} \Y$.

\begin{proposition}\label{pr4a}
The AV-bundle $F^\* \wt{\sT} \wt{\sT} \Y$ is canonically
isomorphic to $(\sT \zt^0_1)^\* \wt{\sT} \Y$, where $\zt^0_1
\colon \sT N \rightarrow N$ is the canonical projection.
    \end{proposition}

    \begin{proof}
An element $\wt{w}$ of $(\sT \zt^0_1)^\* \wt{\sT} \Y$ is
represented by a vector in $\sT \wt{\sT} Y$. If the vector $w
=\wt{\sT} \wt{\sT} (\wt{w}) \in \sT\sT N$ is vertical, then also
$\wt{w} $ is vertical with respect to the canonical projection
$\wt{\zt}_Y \colon\wt{\sT} \Y \rightarrow N$. Since  $\wt{\sT} \Y$
is a vector bundle over $N$, a vertical vector on $\wt{\sT} \Y$
can be identified with an element of $\wt{\sT} \Y$. It follows
that AV-values fibre over $w\in \sT\sT N$ for an affine covector
$\xi^a_F \zs$ can be canonically identified with the fibre of $
\wt{\sT} \Y$ over $\sT \zt^0_1 (w)$.
    \end{proof}

\section{The Euler-Lagrange equation}
Integration by parts in the calculus of variations for curves is
based on the following decomposition of the differential of a
lagrangian $L \colon \sT M \rightarrow \R$

\begin{equation}\label{e-l1}
 (\zt^1_2)^\* \xd L = ((\zt^1_2)^\* \xd L - \xd_\sT (\xi_F \xd L) ) + \xd_\sT (\xi_F \xd L) ,
\end{equation}
where $\zt_2^1 \colon \sT^2 M \rightarrow \sT^1 M = \sT M$ is the
canonical projection. The first component in $\ref{e-l1}$ is a
1-form on $\sT^2 M$, vertical  with respect to projection $\sT^2 M
\rightarrow M$.
  It can be considered as a mapping $\sT^2 M \rightarrow \sT^\* M$

Now, let $\zl \colon \sT M \rightarrow \wt{\sT}$ be an affine
lagrangian, i.e. a section of $\wt{\sT}\zeta$.
    We have the affine version of \ref{e-l1}
\begin{equation}\label{e-l2}
(\zt^1_2)^\*\xd \zl = ((\zt^1_2)^\* \xd \zl \boxminus \xd_\sT
(\xi_F^a \xd \zl) ) \boxplus \xd_\sT^a (\xi_F^a \xd \zl) ,
\end{equation}
    The AV-bundle for the pull-back $(\zt^1_2)^\*\xd \zl$ is (see Section~\ref{pull-back})
    \begin{equation}\label{e-l2a}
(\sT\zt^1_2)^\*(\wt{\sT} \wt{\sT} \Z) = \wt{\sT}((\zt^1_2)^\*
\wt{\sT}\Z) = \wt{\sT}((\sT\zt^0_1)^\* \wt{\sT}\Z) =
\wt{\sT}\wt{\sT} (\zt^0_1)^\* \Z \ .
\end{equation}
We have used that the pull-back commutes with the exterior
derivative and $\zt^1_2$ coincides with $\sT \zt ^0_1$.

    The AV-bundle for $\xi_F^a \xd \zl$ is (see the previous section)
    $$(\sT \zt^0_1)^\* \wt{\sT} \Z = \wt{\sT}((\zt^0_1)^\* \Z) .$$
    This, together with \ref{e-l2a}, implies that the AV-bundle for $((\zt^1_2)^\* \xd \zl \boxminus
\xd^a_\sT (\xi^a_F \xd \zl) )$ is trivial. We can write
$(\zt^1_2)^\* \xd \zl - \xd^a_\sT (\xi^a_F \xd \zl)$, which is
ordinary 1-form on $\sT^2 M$. Since for $\zl = \xi^a_\sT (\xd
\zf)$ we have (arguing as in Proposition~\ref{pr4a}) that
$$ \xi^a_F (\xd \xi^a_\sT (\xd \zf)) = (\zt^0_1)^\* \xd \zf = \xd ((\zt^0_1)^\* \zf)$$
and
$$ \xd^a_\sT (\xi^a_F (\xd \xi^a_\sT (\xd \zf))) = \xd \xi^a_\sT( (\zt^0_1)^\* \xd\zf)
= (\zt^1_2)^* \xd \xi^a_\sT (\xd \zf) \ .$$ It follows that the
first term in the decomposition \ref{e-l2} equals zero. It is
known that the first term in the decomposition \ref{e-l1} is
vertical with respect to the projection $\zt^0_2$, i.e. it can be
considered as a mapping $\cal E \zl \colon \sT^2 M \rightarrow
\sT^\* M$. We conclude, that the same remains valid for affine
lagrangian $\zl$. Similarly, we have that the affine 1-form
$\xi^a_F \xd \zl$ is vertical with respect to the projection
$\zt^0_1 $. It is meaningful because the AV-bundle for $\xi^a_F
\xd \zl$ is the pull-back of $\wt{\sT} \Z$
(Proposition~\ref{pr4a}). Therefore $\xi^a_F \xd \zl$ defines a
mapping $\cal P \zl \colon \sT M \rightarrow \sP \Z$, which is the
affine Legendre map.

 \section{Variational principles} \label{vap}
Variational principles are based on the proper representation of
the differential of the action functional
$$ \cal L \colon \cal M _{[a,b]} \rightarrow \R \colon  \zg \mapsto \int_a^b \zl \circ \dot\zg \ ,$$
where $\cal M _{[a,b]}$  is the space of smooth mappings from the
interval $[a,b]$ to $M$. As we have seen in Section \ref{inhom},
the AV-space for $\cal L$ at $\zg$ is $\Z_{\zg(b)} \boxminus
\Z_{\zg(a)}$.  A  vector  tangent to $\cal M _{[a,b]}$ is a curve
in $\sT M$, i.e a mapping $ w \colon [a,b] \rightarrow \sT M$. The
AV-budle for the differential $\xd \cal L$ at $\zg$ is then
$\wt{\sT}_ {\zg(b)}M \boxminus  \wt{\sT}_ {\zg(a)}M$. A convenient
representation of the differential is suggested by the
decomposition \ref{e-l2}:
$$ \langle \xd \cal L (\zg) , w \rangle = \left(\langle \cal P \zl \circ \dot \zg (b),
w(b)\rangle \boxminus \langle \cal P \zl \circ \dot \zg (a), w(a)
\rangle \right) - \int_a^b \langle \cal E \zl\circ \ddot \zg(t),
w(t)\rangle \ . $$ The above equality suggests that a covector
should be represented by a mapping $f \colon [a,b] \rightarrow
\sT^\* M$ and two affine covectors $p_a $ and $p_b$ in $\sP\Z$.
The variational principle

    $$ \int_a^b \langle \xd \zl, \dot w\rangle  = \langle p_b, w(b)\rangle \boxminus
    \langle p_a, w(a)\rangle - \int_a^b \langle f(t), w(t)\rangle\xd t $$
produces the Euler-Lagrange equation

    $$\cal E \zl \circ \ddot \zg = f$$
 and the momentum-velocity relations
    $$ (\cal P\zl \circ \dot \zg)(a) = p_a, \ \ (\cal P\zl \circ \dot \zg)(b) = p_b \ .$$


\end{document}